\documentstyle[preprint,prb,aps]{revtex}
\begin{document}
\draft
\title{Charged Pseudospin Textures in Double-Layer Quantum Hall
Systems with Spontaneous Interlayer Coherence}

\author{Kun Yang\cite{byline} and A.H. MacDonald}

\address{Physics Department, Indiana University, Bloomington IN 47405}

\date{\today}

\maketitle

\begin{abstract}

We report on a Hartree-Fock approximation study of the meron
pseudospin-texture excitations of the broken-symmetry incompressible
ground states of double-layer quantum Hall states at $\nu =1$.
We have obtained results for meron core energies and core sizes.
We use these to estimate the charge gap which determines the activation
energy for dissipation in the quantum Hall effect.

\end{abstract}


Recent work from several different points of
view\cite{fertig,wenzee,ezawa,gapless,harfok}
has led to the realization that double-layer two-dimensional
electron systems (2DES) in a strong magnetic field can,
under appropriate circumstances, exhibit an unusual broken symmetry in which
interlayer phase coherence develops in the absence of interlayer tunneling.
This broken symmetry has useful analogies with the spin-ferromagnetism
which occurs\cite{sondhi} in a single-layer 2DES in the quantum
Hall regime, for example, at odd integral Landau
level filling factors, $\nu$.
In the pseudospin representation commonly used for the (spin-polarized)
double-layer system, spontaneous interlayer phase coherence
is equivalent to the
spontaneously broken U(1) symmetry of an
easy-plane ferromagnet.\cite{letter,dl1}  In the limit of zero separation
$d$ between the layers, the (spin-polarized)
double-layer system with no tunneling is equivalent to
the single-layer system with no Zeeman coupling; in this limit the
ferromagnetism of the pseudospin representation for the double-layer
2DES becomes isotropic.  Interesting physical
consequences follow from spontaneous interlayer
phase coherence, including the occurrence of a
finite temperature Kosterlitz-Thouless phase
transition\cite{wenzee,letter,dl1}, superfluid-like behavior in the
pseudospin channel,\cite{letter,dl1,ezawa} and a commensurate-incommensurate
phase transition driven by an in-plane component of the external
magnetic field in the presence of weak interlayer
tunneling.\cite{murphy,letter,dl2}
Argueably, the most remarkable property associated with the broken
symmetry is the occurrence of highly collective,
topologically nontrivial
pseudospin texture excitations which carry a physical charge equal (in units
of the electron charge) to
the topological charge of the texture.\cite{letter,dl1}
These charged pseudospin textures are analogous to the skyrmion
charged spin textures of single-layer systems\cite{sondhi,leekane,fertig94}
whose presence has recently been detected experimentally.\cite{barrett}
In this paper we report on a Hartree-Fock calculation
which estimates the energy of these excitations for double-layer 2DES's
at $\nu =1$.

The fractional quantum Hall effect, an anomaly in the transport
properties of 2DES's in a strong magnetic field\cite{myreview},
occurs when the system has a charge gap, {\it i.e.,} a
discontinuity in the chemical potential, at a critical density
which depends on magnetic field.   (It is the magnetic field dependence
of the critical density which leads to the
charge carried by the spin and pseudospin textures.\cite{dl1})
At the critical density the dissipative conductivity, $\sigma_{xx}$,
has an activated temperature dependence with an activation energy
which, for the case of interest here, is half the charge gap.
In single-layer systems, it is known\cite{sondhi,fertig94} that the lowest
energy charged excitations at $\nu=1$ (but not at larger odd integral values
of $\nu$) are skyrmion spin-textures.
It follows from continuity that in spin-polarized double-layer systems
the charge gap must also be given by the energy of its charged spin textures,
at least if the layer separation, $d$, is not too large.
In the double-layer case, where the broken symmetry is planar, vortices might
be expected to play an important role.  Indeed, it has been
proposed\cite{letter,dl1} that the charged pseudospin textures of
double-layer systems consist of meron-antimeron pairs.  Merons are
vortex-like pseudospin textures where the pseudospin tilts out of the
easy plane in the vortex core; neither the core size nor the core energy
are fixed by the parameters which characterize the system at long length
scales\cite{sondhi,dl1} and both must be
determined by microscopic calculations.
Merons have an infinite energy and
carry\cite{letter,dl1} charge $ \pm e/2$, but meron pairs
with like charges and opposite vorticities carry charge $\pm e$ and have
a finite energy.  In this paper we evaluate the meron-core energy and size
using a Hartree-Fock approximation and use the results to estimate
the energy of the charge $\pm e$ excitations of double-layer systems and
to limit the range of validity of the meron pair description of these
excitations.

In a recent paper,\cite{dl1} Moon {\em et al.} introduced
single-Slater-determinant microscopic
wave functions for both skyrmion and meron states near $\nu = 1$,
for which the spin and pseudospin textures have the appropriate topological
charges.  The form of these wave functions provides an understanding
of the connection between topological charge and physical charge from
a microscopic physics point of view, as we discuss below, but
the wavefunctions were not energetically optimized.
For the single-layer case, Fertig {\em et al.}\cite{fertig94}
were able to find the optimal single-Slater-determinant wavefunctions
for a skyrmion spin texture as a function of the strength of the
Zeeman coupling by solving a set of Hartree-Fock equations.
In this paper we generalize the Hartree-Fock calculations to the
case of double-layer systems and find the optimal single-Slater-determinant
approximation to the meron state as a function of $d$.
The technical details of our calculation are very
similar to those of Fertig {\em et al.}.  However, meron states
are quite different from skyrmion states and have richer physics
as we demonstrate below.

We assume that the Landau level spacing is so large that all
electron stay in the lowest Landau level (LLL), hence we only need to keep
one body states in the LLL.  To take advantage of the circular symmetry
of the meron state, we use the symmetric gauge in
which the one body wave functions in the LLL have the form
$\phi_m({\bf r})= (2\pi 2^m m!)^{-1/2} z^m\exp(-|z|^2/4)$,
where $z=x+iy$ is the complex coordinate, $m$ is the angular momentum
quantum number and lengths are expressed in units of the
magnetic length. ($\ell \equiv \hbar c/eB$ where
$B$ is the magnetic field strength). In the limit $d=0$ (which is
equivalent to the real spin case), inter- and intralayer
Coulomb interaction have the same form and the system is invariant under
pseudospin rotation.\cite{letter,dl1}  In the Hartree-Fock approximation,
the ground state\cite{fertig,harfok} at $\nu =1$
in the standard pseudospin representation is completely
polarized along an arbitrary direction in the
$\hat x - \hat y $ plane:
\begin{equation}
|\Psi_0\rangle=\prod_{m=0}^{N-1}({1\over \sqrt{2}}c^{\dagger}_{m\uparrow}
+{e^{i \phi} \over \sqrt{2}}c^{\dagger}_{m\downarrow})\vert 0\rangle.
\label{gs}
\end{equation}
Here $|0\rangle$ is the vacuum state,
$c^{\dagger}_{m\uparrow (\downarrow)}$ creates an
electron in the $\phi_m$ state in the upper (lower) layer, $N$ is
the number of electrons, and $\phi$ is an arbitrary phase.
(In the following we choose $\phi =0$ so
that the pseudospin is oriented in the $\hat x$ direction in the ground
state.)
Note that the orbitals are peaked farther from the origin as the angular
momentum increases.  For $d=0$, the state in Eq.~(\ref{gs})
is the exact ground state\cite{dl1} and we expect it to remain
accurate as long as $d$ is not too large.

The single-Slater-determinant states for
merons centered at the origin with vorticity $+1$
and with charge $ -e/2$ and $ + e/2 $ have the form\cite{dl1}:
\begin{equation}
\vert\Psi_{+1,-{1\over 2}}\rangle
=\prod_{m=0}^{N-1}(u_mc^{\dagger}_{m\uparrow}
+v_mc^{\dagger}_{m+1\downarrow})\vert 0\rangle,
\label{hole}
\end{equation}
\begin{equation}
\vert\Psi_{+1,+{1\over 2}}\rangle
=c^{\dagger}_{0\downarrow}\prod_{m=0}^{N-2}
(-v_mc^{\dagger}_{m\uparrow}
+u_mc^{\dagger}_{m+1\downarrow})\vert 0\rangle.
\label{particle}
\end{equation}
Here
$u_m$ and $v_m$ are variational parameters which, for
$N \to \infty $, have the same
values in both wavefunctions because of the particle-hole symmetry
of the Hamiltonian\cite{fertig}, and which satisfy the constraint
$|u_m|^2+|v_m|^2=1$
%\begin{equation}
%|u_m|^2+|v_m|^2=1
%\end{equation}
so that the wavefunction is normalized.
The easy-plane anisotropy of the double-layer system requires that
\begin{equation}
\lim_{m\rightarrow\infty}{|u_m|\over|v_m|}=1,
\end{equation}
so that at the charge density is the same for both layers far
enough from the meron center and the
electrostatic energy cost of the meron texture is finite.\cite{letter,dl1}
The charges of the two states are $\pm e/2$ as can be verified
by evaluating the
differences in mean occupation numbers between the meron states
and the ground state.

In this paper we assume, without loss of generality,
that $u_m$ and $v_m$ are real.  Far from the origin, single-particle
orbitals with adjacent angular momenta have nearly identical
magnitudes near their peaks where they satisfy
$\phi_{m+1}({\bf r}) /\phi_{m}({\bf r}) \approx \exp(i \phi) $ where $\phi$ is
the angular coordinate.  It follows that for large $m$,
the single-particle state
in the Slater determinant is effectively proportional to
the pseudospinor whose
azimuthal angle is equal to the angular coordinate:
\begin{equation}
\chi(\phi)={1\over \sqrt{2}}\left(\begin{array}{c}
\cos(\theta/2)\\
\sin(\theta/2) \exp(i \phi)\end{array}\right).
\end{equation}
(We have imposed the normalization constraint by setting $u_m =
\cos(\theta_m/2)$ and $v_m = \sin(\theta_m/2)$. $\theta$ is the
polar angle for the pseudospin contribution from orbital $m$.)
At large $m$, $\theta = \pi/2$, the pseudospinor is confined to the
$\hat x- \hat y$ plane and has vorticity $+1$.
At the center of
the meron only $\phi_{m=0}({\bf r}) \ne 0$ so that the pseudospinor
points up for the charge $-e/2$ meron and down for the charge $e/2$ meron.
Meron states with opposite pseudospin vorticity
can be generated by interchanging the roles of the two layers.

The Hartree-Fock equations we solve determine the $u_m$ and $v_m$
which minimize the expectation value of the Hamiltonian in states of the
form of Eq.~(\ref{hole}) and Eq.~(\ref{particle}).  We perform our
calculation at finite $N$ (up to $N=60$)
and to mitigate finite-size effects
we include in the Hamiltonian the external potential
from a fixed background which neutralizes the ground state charge density.
The full Hamiltonian of the system is
\begin{eqnarray}
H={1\over 2}\sum_{m_1m_2m_3m_4}\sum_{\sigma_1\sigma_2}
{V^{\sigma_1\sigma_2}_{m_1m_2m_3m_4}c^{\dagger}_{m_1\sigma_1}
c^{\dagger}_{m_2\sigma_2}
c_{m_4\sigma_2}c_{m_3\sigma_1}}\nonumber\\
-\sum_{m}\sum_{\sigma}{c_{m\sigma}^{\dagger}c_{m\sigma}[\sum_{n=0}^{N-1}
{{1\over 2}(V^A_{mnmn}+V^E_{mnmn})}]},
\label{H}
\end{eqnarray}
where $\sigma_i = \uparrow$ or $ \downarrow$;
$V^{\uparrow\uparrow}=V^{\downarrow\downarrow}=V^A$ is the intralayer
interaction and $V^{\uparrow\downarrow}=V^{\downarrow\uparrow}$ is the
interlayer interaction.   For the sake of definiteness we have taken the
electron layers to have negligible thickness so that
\begin{equation}
V^A_{m_1m_2m_3m_4}=\int{d{\bf r_1}d{\bf r_2}}\phi^*_{m_1}({\bf r_1})
\phi^*_{m_2}({\bf r_2}){e^2\over |{\bf r_1}-{\bf r_2}|}
\phi_{m_3}({\bf r_1})\phi_{m_4}({\bf r_2}),
\label{eq:intra}
\end{equation}
and
\begin{equation}
V^E_{m_1m_2m_3m_4}=\int{d{\bf r_1}d{\bf r_2}}\phi^*_{m_1}({\bf r_1})
\phi^*_{m_2}({\bf r_2}){e^2\over \sqrt{|{\bf r_1}-{\bf r_2}|^2+d^2}}
\phi_{m_3}({\bf r_1})\phi_{m_4}({\bf r_2}).
\label{eq:inter}
\end{equation}
The one body term in (\ref{H}) is due to the interaction
of the electrons with the neutralizing background charge; it reduces to
an irrelevant constant for $N \to \infty$ and, although we include it
in our numerical calculations, we omit it to free the discussion below from
unnecessary clutter.

An elementary but somewhat tedious calculation gives the following result
for the expectation value of the Hamiltonian in the charge $-e/2$
meron state:
\begin{eqnarray}
\langle \Psi_{+1,-{1\over 2}} \vert H \vert\Psi_{+1,-{1\over 2}}\rangle
&=&
{1 \over 2} \sum_{m,n=0}^{N-1}
\big[ u_m^2 u_n^2 (V^A_{mnmn}-V^A_{mnnm})\nonumber\\
&&+v_m^2 v_n^2
(V^A_{m+1,n+1,m+1,n+1} -V^A_{m+1,n+1,n+1,m+1})\nonumber\\
&&+2 u_m^2 v_n^2 V^E_{m,n+1,m,n+1} -2 u_m v_m u_n v_n
V^E_{m,n+1,n,m+1} \big].
\label{eq:totaleng}
\end{eqnarray}
Minimizing Eq.~(\ref{eq:totaleng}) with respect to $\theta_m$
gives the following equation which must be solved self-consistently
for the pseudospinor polar angles,
\begin{equation}
\tan(\theta_m) = {A_m/B_m},
\label{eq:selfcons}
\end{equation}
where
\begin{equation}
A_m=-2 \sum_{n=0}^{N-1} V^E_{m,n+1,n,m+1} \sin (\theta_n)
\end{equation}
and
\begin{eqnarray}
B_m&=&
\sum_{n=0}^{N-1} [V^A_{mnmn}-V^A_{mnnm}-V^A_{m+1,n+1,m+1,n+1}
+V^A_{m+1,n+1,n+1,m+1}\nonumber\\
&&+V^E_{m,n+1,m,n+1}-V^E_{n,m+1,n,m+1} +\cos (\theta_n)
(V^A_{mnmn}+V^A_{m+1,n+1,m+1,n+1}\nonumber\\
&&-V^A_{mnnm}-V^A_{m+1,n+1,n+1,m+1}
-V^E_{m,n+1,m,n+1}-V^E_{n,m+1,n,m+1})].
\end{eqnarray}
Similar equations can be obtained for merons with charge $+e/2$.
The various terms in $A_m$ and $B_m$ can be understood
rather simply.  For large $m$ the matrix elements are
insensitive to unit shifts in the angular momentum indices so that the
constant terms in the denominator vanish rather quickly.  These
unit angular momentum shifts are important in determining the
the structure of the meron core but we will temporarily ignore them
for the following qualitative discussion.  The term
proportional to $\cos (\theta_n)$ in the denominator is the sum of
intralayer exchange matrix elements and the difference of direct
matrix elements
for intralayer and interlayer interactions.
If $V^A$ were equal to $V^E$ Eq.~(\ref{eq:selfcons}) would be
satisfied for any constant $\theta_n$, corresponding to the pseudospin
isotropy at $d=0$.   The difference between intralayer and interlayer
matrix elements which appears in the denominator of
Eq.~(\ref{eq:selfcons}) gradually forces $\theta_m$ to $\pi/2$ as we move
away from the meron core.

The most important result of our calculation is shown in Fig.~[\ref{fig1}]
where we plot our estimates of the chemical potentials at densities just
larger and just smaller than the critical density at which the incompressible
state occurs, $\mu^{+}_{MP}$ and $\mu^{-}_{MP}$ respectively.
These chemical potentials were obtained from the optimized meron state
energies by the following procedure.  For each value of $d$ the difference
between the meron state energy and the ground state energy, $E_m^{\pm}$
was evaluated at a series of $N$ values.  The meron core energies,
$E_{mc}^{\pm}$ were extracted by fitting our results to the expected form
\begin{equation}
E_m^{\pm}=E_{mc}^{\pm}+\pi\rho_E\ln (R/R_{mc}) ,
\end{equation}
where $R=\sqrt{2(N-1)}\ell$ is the radius of the system,
$\rho_E$ is the Hartree-Fock approximation for the
in-plane pseudospin stiffness\cite{dl1},
\begin{equation}
\rho_E={1\over 32\pi^2}\int_0^{\infty}{dk}k^3V^E(k)e^{-k^2/2},
\end{equation}
$V_E(k)$ is the Fourier transform of the interlayer interaction.
The meron core radius $R_{mc}$ (which is identical for merons with
charges $\pm e/2$)
was defined by $m^z(R_{mc})=4\pi\ell^2\langle
S^z(R_{mc})\rangle=0.1$, where $S^z({\bf r})$
is the $z$ component of the pseudo spin density operator.\cite{dl1}
The energy of an individual meron is infinite because of the
large distance contribution to the gradient energy, but the energy
of a meron pair of opposite vorticity is finite.  The optimal
separation\cite{dl1}, $R^*$, for the meron pair is determined by minimizing
the sum of the Coulomb energy and the gradient energy.  For $ R >> R_{mc}$
we find\cite{dl1} that $R^*=e^2/8 \pi \rho_E$ and $E_{MP}^{\pm}
= 2 E_{mc}^{\pm} + 2 \pi \rho_E ( 1 + \ln (R^*/R_{mc}))$.
In Fig.~[\ref{fig2}] we plot $R^*$ and
$R_{mc}$ as a function of $d$; the meron pair picture of the charge
excitations is valid only when $R > R_{mc}$ which is satisfied
for $d/\ell > 0.6$.  Results at smaller values of $d/\ell$ in
Fig.~[\ref{fig1}] and
Fig.~[\ref{fig2}] are plotted as dotted lines.  $E_{MP}^{\pm}$ is the energy to
make a meron pair at fixed total electron number by adding or removing
charge from the edge of the system to compensate for the charge of a meron;
to add or remove an electron it is necessary to change the electron number
in the incompressible state.  It follows\cite{myreview,qppaper,unpub} that
\begin{equation}
\mu^{\pm}_{MP} = \pm E_{MP}^{\pm} + \epsilon_0(d)
\label{eq:chempot}
\end{equation}
where $\epsilon_0(d)$ is the energy per electron in the incompressible state.
We remark that as a consequence of a particle-hole symmetry which
applies in the thermodynamic limit $\mu^{+}_{MP} + \mu^{-}_{MP} =
2 \epsilon_0(0)$. We have used this identity to check the reliability of
our numerical procedures, particularly those involved in
the extraction of the meron core energy from the size dependence of the
meron energy.  For $d/\ell > \approx 1.2$ we do not believe that our
estimates are reliable; results in this range of $d / \ell$ are
also shown as dotted lines in Fig.~[\ref{fig1}] and Fig.~[\ref{fig2}].  The
problem at
these values of $d/\ell$ is connected with the fact that the Hartree-Fock
ground state at $\nu=1$ is\cite{nosdw} a pseudospin-density-wave\cite{harfok}
for $d/\ell > \approx 1.2$. Within the range of $d$ that the meron pair
picture of charge excitations is valid, the meron-pair states are clearly
energetically more favorable than naive Hartree-Fock single particle states
(see Fig.~[\ref{fig1}])
and our results agree with exact diagonalization studies\cite{dl1}
qualitatively.

We thank L. Brey, R. C\^{o}t\'{e}, H.A. Fertig,
S.M. Girvin, F.D.M. Haldane, K. Moon, N. Read, and S. Sondhi  for helpful
discussions. This work was supported by NSF grants DMR-9416906 and
DMR-9224077.

\begin{figure}
\caption{Estimates of the chemical potentials
(in units of $e^2/(\epsilon\ell)$)
just above ($\mu^+_{MP}$) and below ($\mu^-_{MP}$)
$\nu=1$ as a function of $d/\ell$,
assuming that the meron pair state is a good description of the
charged excitations. The dotted portions of the lines are
for the ranges of $d$ were the meron pair picture is
invalid and our estimate is unreliable. The
difference between $\mu_{MP}^+$ and $\mu^-_{MP}$ is the
charge gap of the system at $\nu=1$. For comparison we also plot the
chemical potentials for the Hartree-Fock single particle excitations
(dashed lines) which cost more energy than the meron pairs.}
\label{fig1}
\end{figure}

\begin{figure}
\caption{Estimates of the optimal meron core size $R_{mc}$ and optimal
separation $R^*$ between merons
in a charged meron pair state as a function of $d/\ell$.
The dotted portions of these curves are in ranges of layer separation where
the meron pair picture is not valid.}
\label{fig2}
\end{figure}

\end{document}